\documentstyle[12pt,epsf,epsfig,wrapfig]{article}
\newcommand{\be}{\begin{equation}}
\newcommand{\ee}{\end{equation}}
\newcommand{\ba}{\begin{eqnarray}}
\newcommand{\ea}{\end{eqnarray}}
\newcommand{\nn}{\nonumber}

\def\mev{~{\rm MeV}}
\def\gev{~{\rm GeV}}

\newcommand{\gsim}{\raisebox{-4pt}{$\,\stackrel{\textstyle
                                                         >}{\sim}\,$}}
\newcommand{\req}[1]{(\ref{#1})}
\def\={\,=\,}
\newcommand{\ci}[1]{\cite{#1}}

\begin{document}

\begin{center}
{\bfseries GENERALIZED PARTON DISTRIBUTIONS\\[0.2em]
        FROM NUCLEON FORM FACTORS}

\vskip 5mm
P.\ Kroll

\vskip 5mm 
{\small {\it Fachbereich Physik, Universit\"at Wuppertal, Germany}\\
{\it kroll@physik.uni-wuppertal.de}}
\end{center}

\vskip 5mm
\begin{abstract} 
Results from a recent analysis of the zero-skewness generalized parton
distributions (GPDs) for valence quarks are reviewed. The analysis
bases on a physically motivated parameterization of the GPDs with a
few free parameters adjusted to the nucleon form factor data. 
The Fourier transforms of the GPDs representing quark densities in
the impact parameter plane, as well as moments of the GPDs are also 
discussed. The $1/x$ moments in particular form the soft physics input
to Compton scattering off protons within the handbag approach. The 
Compton cross section evaluated from this information is found to be
in good agreement with experiment.\\

Talk presented at the XI workshop on High Energy Spin Physics (SPIN
2005), JINR Dubna, September 2005
\end{abstract}

\vskip 6mm
\section{Introduction}
In recent years we have learned how to deal with hard exclusive
reactions within  QCD. In analogy to hard inclusive reactions the
process amplitudes factorize in partonic subprocess amplitudes, 
calculable in perturbation theory, and in GPDs which parameterize soft
hadronic matrix elements. In some cases rigorous proofs of
factorization exist. For other processes factorization is shown to
hold in certain limits, under certain assumptions or is just a 
hypothesis. The GPDs which are defined by Fourier transforms of
bilocal proton matrix elements of quark field operators \ci{mue1994}, 
describe the emission and reabsorption of partons by the proton. For 
equal helicities of the emitted and reabsorbed parton the structure of
the nucleon is described by four GPDs, termed $H$, $\widetilde{H}$, $E$ 
and $\widetilde{E}$, for each quark flavor and the gluons. The GPDs
are functions of three variables, the momentum transfer from the
initial to the final proton, $t$, a momentum fraction $x$ and the 
skewness, $\xi$. The latter variable is related to the difference of 
the individual momentum fractions the emitted and reabsorbed partons 
carry. The GPDs are subject to evolution and, hence, depend on the 
factorization scale $\mu$, too. They satisfy the reduction formulas
\be
H^q(x,\xi=0,t=0,\mu)\= q(x,\mu)\,, \qquad  
                        \widetilde{H}^q(x,\xi=0,t=0,\mu)\= \Delta q(x,\mu)\,,  
\ee
i.e.\ in the forward limit 
$H$ and $\widetilde{H}$ reduce to the usual unpolarized 
and polarized parton distributions (PDFs), respectively. Another 
property of the GPDs is the polynomiality which comes about as a 
consequence of Lorentz covariance  
\be
\int^1_{-1}\, dx\, x^{m-1}\, H^q(x,\xi,t,\mu) \= \sum^{[m/2]}_{i=0}\,
h^q_{m,i}(t,\mu)\,\xi^i\,,
\label{pol}
\ee 
where $[m/2]$ denotes the largest integer smaller than or equal to
$m/2$. Eq.\ \req{pol} holds analogously for the other GPDs and, for
$m=1$, implies sum rules for the form factors of the nucleon, e.g.\
\be
F^q_1(t) \equiv h^q_{1,0}(t) \= \int^1_{-1}\, dx H^q(x,\xi,t,\mu)\,.
\label{sumrule}
\ee
Reinterpreting as usual a parton with a negative momentum fraction $x$
as an antiparton with positive $x$ ($H^{\bar{q}}(x)=-H^q(-x)$), one 
becomes aware that only the difference of the contributions from
quarks and antiquarks of given flavor contribute to the sum rules. 
Introducing the combination
\be
H^q_v(x,\xi,t,\mu) \=   H^q(x,\xi,t,\mu) -  H^{\bar{q}}(x,\xi,t,\mu)\,,
\ee
for positive $x$ which, in the forward limit, reduces to the usual
valence quark density $q_v(x)=q(x)-\bar{q}(x)$, one finds for the
Dirac form factor the representation
\be
F_1^{p(n)}(t)\= e_{u(d)} \int_0^1\, dx\, H_v^u(x,\xi,t,\mu) + e_{d(u)}
\int_0^1\, dx\, H_v^d(x,\xi,t,\mu)\,.
\label{pn-sr}
\ee     
Here, $e_q$ is the charge of the quark $q$ in units of the positron
charge. Possible contributions from other flavors, $s-\bar{s}$ or 
$c -\bar{c}$, are neglected in the sum rule \req{pn-sr}.
A representation analogous to \req{pn-sr} holds for the 
Pauli form factor with $H$ being replaced by $E$.

The isovector axial-vector form factor satisfies the sum rule 
($\widetilde{H}^{\bar{q}}(x)=\widetilde{H}^{q}(-x)$)
\be
F_A(t) \= \int_0^1 dx\, \left\{\widetilde{H}_v^u(x,\xi,t,\mu) -
                                  \widetilde{H}_v^d(x,\xi,t,\mu)
                   + 2 \, \Big[\widetilde{H}^{\bar{u}}(x,\xi,t,\mu) -
                       \widetilde{H}^{\bar{d}}(x,\xi,t,\mu)\Big]\right\}\,,
\label{axial-sr}
\ee
At least for small $t$ the magnitude of the second integral in Eq.\ 
\req{axial-sr} reflects the size of the flavor non-singlet combination 
$\Delta \bar{u}(x) - \Delta \bar{d}(x)$ of forward densities. This 
difference is poorly known, and at present there is no experimental 
evidence that it might be large \ci{HERMES}. For instance, in the
analysis of the polarized PDFs performed by Bl\"umlein and 
B\"ottcher~Ref.\ \ci{BB} it is zero. In a perhaps crude approximation 
the second term in \req{axial-sr} can be neglected. These simplifications 
to the sum rules \req{pn-sr} and \req{axial-sr} do not imply that the 
nucleon is assumed to consist solely of valence quarks. Sea quarks are 
there but the virtual photon that probes the quark content of the
nucleon, sees only the differences between quark and antiquark 
distributions which are likely small.
\section{A determination of the zero-skewness GPDs}
Since the GPDs cannot be calculated from QCD with a sufficient degree of
accuracy at present we have either to rely on models or to extract
them from experiment as it has been done for the PDFs, see for
instance Refs.\ \ci{BB,CTEQ}. 
A first attempt to extract the GPDs phenomenologically has been
carried through in Ref.\ \ci{DFJK4} (for a similar analysis see Ref.\
\ci{guidal}). The idea is to exploit the sum rules \req{pn-sr} and
\req{axial-sr} and to determine the GPDs from the data~\ci{brash} on
the nucleon form factors, $F_1$ and $F_2$ for proton and neutron as well as
$F_A$. Since the sum rules represent only the first moments of the GPDs 
this task is an ill-posed problem in a strict mathematical sense. 
Infinitely many moments are needed to deduce the integrand, i.e.\ the
GPDs, unambiguously  from an integral. However, from phenomenological 
experience with particle physics one expects the GPDs to be smooth
functions and a small number of moments will likely suffice to fix the
GPDs. The extreme - and at present the only feasible - point of view
is that the lowest moment of a GPD alone is sufficient \ci{DFJK4}. 
Indeed, using recent results on PDFs~\ci{BB,CTEQ} and form factor 
data~\ci{brash} in combination with physically motivated 
parameterizations of the GPDs, one can carry through this analysis. 
Needless to say that this method while phenomenologically successful  
as will be discussed below, does not lead to unique results. Other 
parameterizations which may imply different physics, cannot be
excluded at the present stage of the art. 

The sum rules \req{pn-sr}, \req{axial-sr} are valid at all $\xi$ but
guessing a plausible parameterization of a function of three variables
is a nearly hopeless task. Choosing the special value $\xi=0$ for which
the emitted and reabsorbed partons carry the same momentum fractions, 
has many advantages. One exclusively works in the so-called DGLAP
region where $\xi\leq x$. In this region parton ideas apply and the Fourier
transform of a GPD with respect to the momentum transfer 
${\bf \Delta}$ ($\Delta^2=-t$) has a density interpretation in the
impact parameter plane~\ci{burk}. Moreover, as shown in 
Refs.\ \ci{DFJK1,rad98}, wide-angle exclusive
processes are controlled by generalized form factors that represent
$1/x$ moments of zero-skewness GPDs. The parameterization that is
exploited in Ref.\ \ci{DFJK4} combines the usual PDFs with an exponential 
$t$ dependence (the arguments $\xi=0$ and $\mu$ are omitted in the 
following)
\be
H_v^q(x,t) \= q_v(x) \exp[tf_q(x)]\,,
\label{ansatz}
\ee
where the profile function reads 
\be
f_q(x) \= \big[\alpha'\, \log(1/x) + B_q\big](1-x)^{n+1} + A_q x (1-x)^n\,.
\label{profile}
\ee 
This ansatz is motivated by the expected Regge behavior at low $-t$
and low $x$ \ci{arbarbanel} ($\alpha'$ is the Regge slope for
which the value $0.9\,\gev^2$ is imposed). For large $-t$ and large
$x$, on the other hand, one expects a behavior like $f_q \sim 1-x$
from overlaps of light-cone wavefunctions~\ci{DFJK1,DFJK3}. The ansatz 
\req{ansatz}, \req{profile} interpolates between the two limits 
smoothly~\footnote{
The parameter $B_q$ is not needed if $\alpha'$ is freed. A fit to the
data of about the same quality and with practically the same results
for the GPDs is obtained for $\alpha^\prime\simeq 1.4$.} 
and allows for a  stronger suppression of $f_q$ in the limit $x\to 1$.  
It matches the following criteria for a good parameterization:
simplicity, consistency with theoretical and phenomenological
constraints, stability with respect to variations of PDFs and
stability under evolution (scale dependence of the GPDs can be absorbed 
into parameters).

\begin{figure}
\begin{center}
\includegraphics[width=.30\textwidth, height=.33\textwidth,
  bb=77 448 399 786,clip=true] {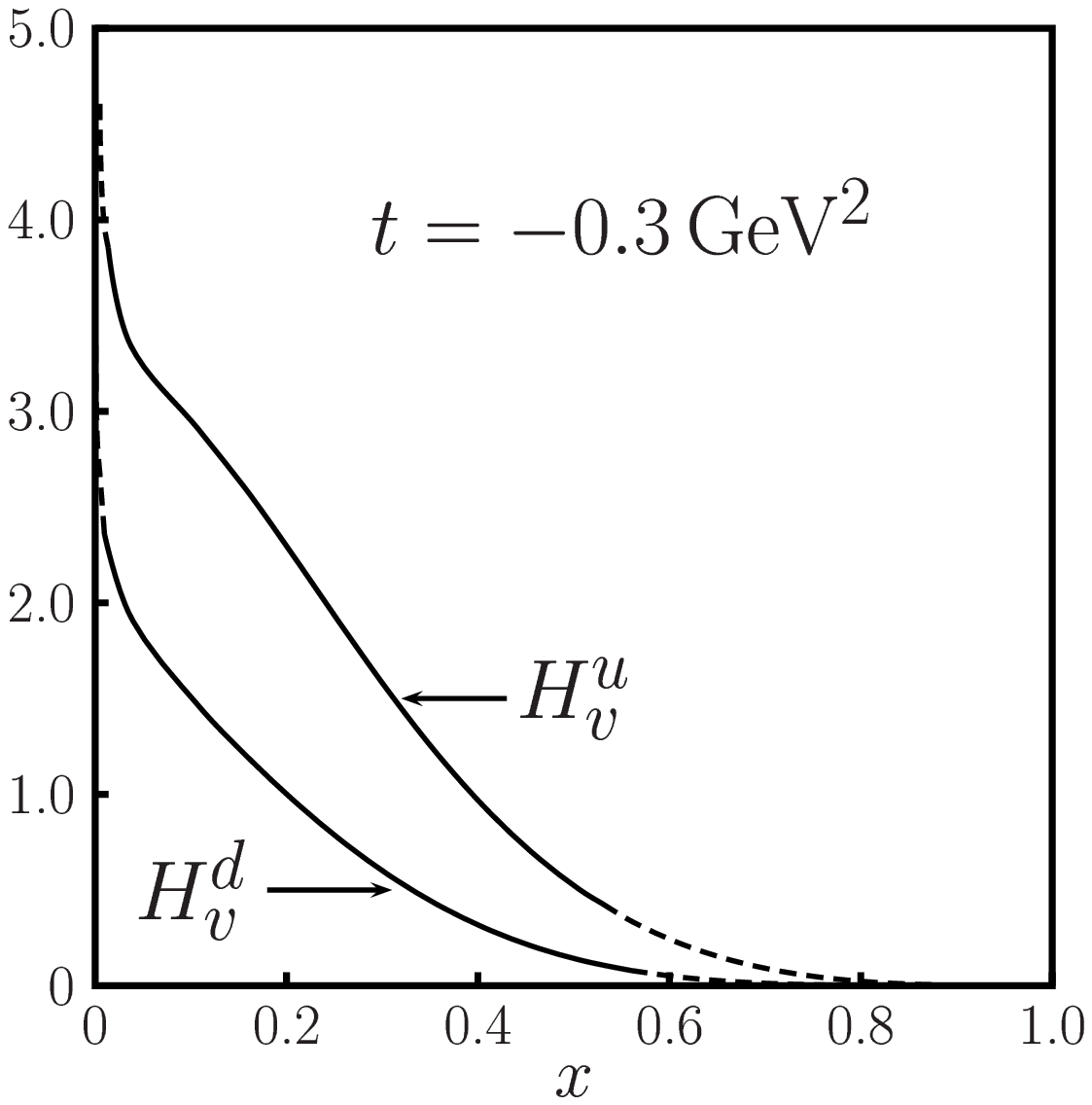}
\hspace{1em} 
\includegraphics[width=.30\textwidth, height=.33\textwidth,
  bb=76 294 400 628,clip=true] {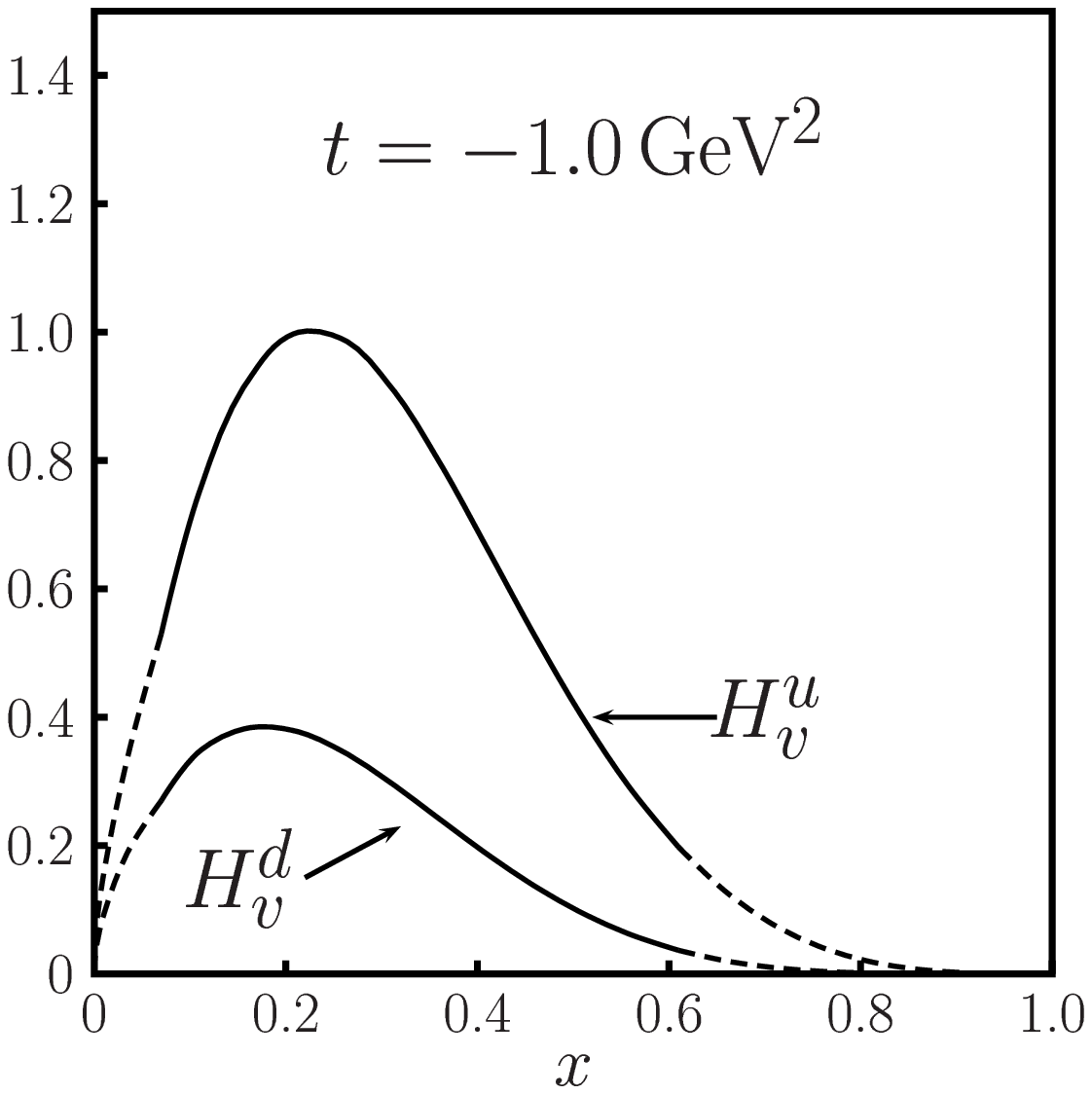}
\hspace{1em} 
\includegraphics[width=.30\textwidth, height=.33\textwidth,
  bb=66 479 400 826,clip=true] {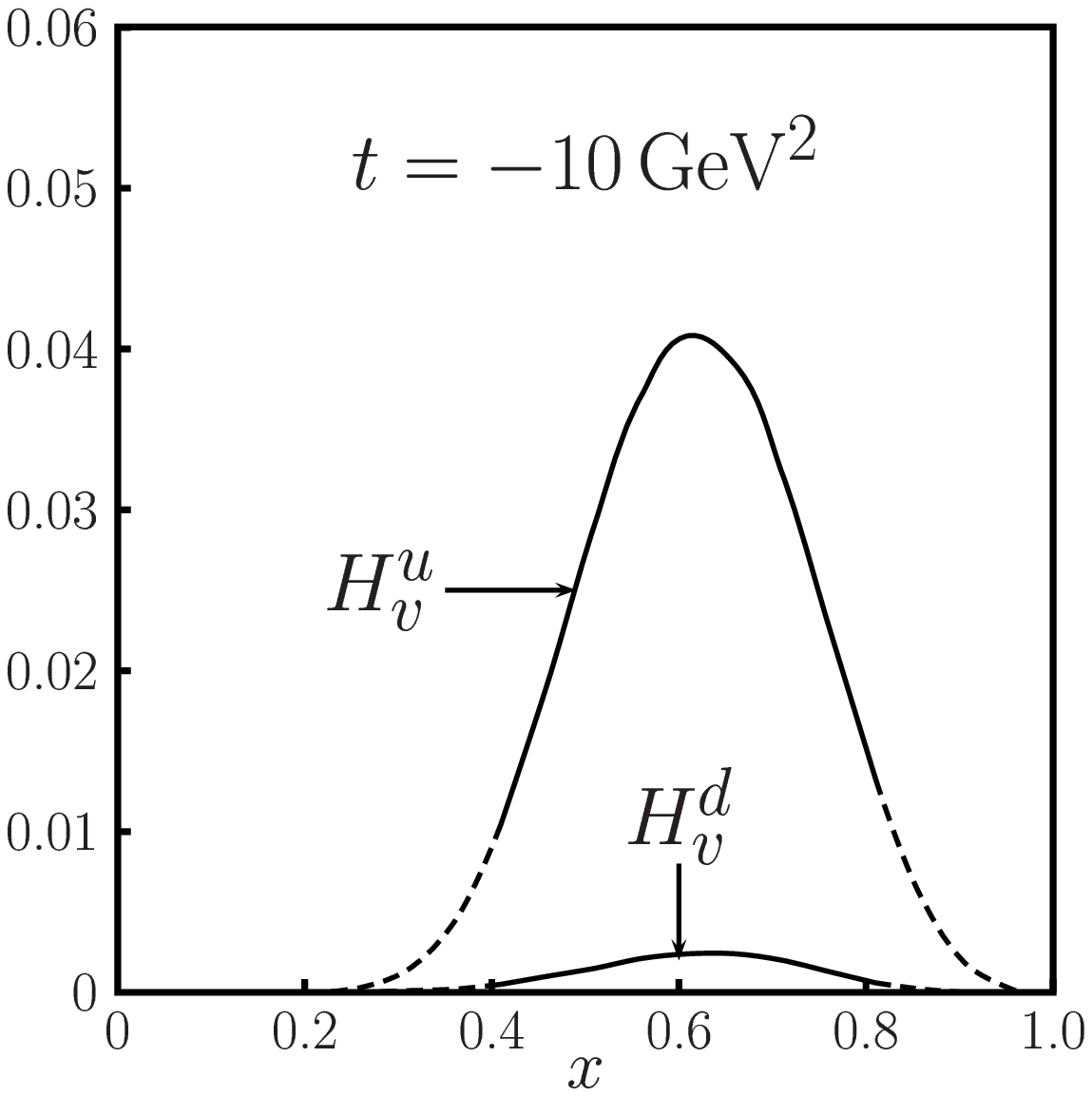}
\end{center}
\vspace*{-1.6em}
\caption{\label{fig:Hgpd} Results for $H_v^q(x,t)$ at the scale 
$\mu=2\,\gev$ and for $n=2$ obtained in \cite{DFJK4}. In each of the
  regions indicated by dashed lines, $5\%$ of the total value of the 
sum rule \req{pn-sr} are accumulated.}
\label{fig:gpdH}
\end{figure}

Using the CTEQ PDFs \ci{CTEQ}, one obtains a reasonable fit of the
ansatz \req{ansatz}, \req{profile} to the data on the Dirac form
factor (from $-t=0$ up to $\simeq 30\, \gev^2$) with the
parameters
\ba
B_u&=&B_d\=(0.59\pm 0.03)\,\gev^{-2}\,, \nn\\
 A_u&=&(1.22\pm 0.020)\,\gev^{-2}\,, \quad A_d= (2.59\pm 0.29)\,\gev^{-2}\,, 
\ea
quoted for the case $n=2$ and at a scale of $\mu=2\,\gev$. In 
Fig.\ \ref{fig:gpdH} the results for $H^q_v$ are shown at three values
of $t$. While at small $-t$ the behavior of the GPD still reflects
that of the parton densities it exhibits a pronounced maximum at
larger values of $-t$. The maximum becomes more pronounced with
increasing $-t$ and its position moves towards $x=1$. In other words
only a limited range of $x$ contributes to the form factor
substantially.
The quality of the fit is very similar in both the cases, $n=1$ and 2;
the results for the GPDs agree well with
each other. Substantial differences between the two results only occur
for very low and very large values of $x$, i.e.\ in the regions which are
nearly insensitive to the present form factor data. It is the physical
interpretation of the results which favours the fit with $n=2$. Indeed
the average distance between the struck quark and the cluster of spectators
becomes unphysical large for $x\to 1$ in the case $n=1$; it grows like
$\sim (1-x)^{-1}$ while, for $n=2$, it tends towards a constant value of
about 0.5 fm \ci{DFJK4}. 

The analysis of the axial and Pauli form factors, with
parameterizations analogous to Eqs.\ \req{ansatz}, \req{profile}, provides the
GPDs $\widetilde{H}$ and $E$. In general they behave similar to $H$. Noteworthy
differences are the opposite signs of $\widetilde{H}^u_v$ ($E^u_v$) and
$\widetilde{H}^d_v$ ($E^d_v$) and the approximately same magnitude of 
$E^u_v$ and $E^d_v$ at least for not too large values of $-t$. For $H^q_v$
and $\widetilde{H}^q_v$, on the other hand, the $d$-quark contributions
are substantially smaller in magnitude than the $u$-quark ones, see
Fig.~\ref{fig:gpdH}. Since there is no data available for the
pseudoscalar form factor of the nucleon the GPD $\widetilde{E}$ cannot
be determined this way. 
\section{Moments of the GPDs}
\begin{figure}
\begin{center}
\includegraphics[height=.42\textwidth,
 bb = 48 296 444 655,clip=true]{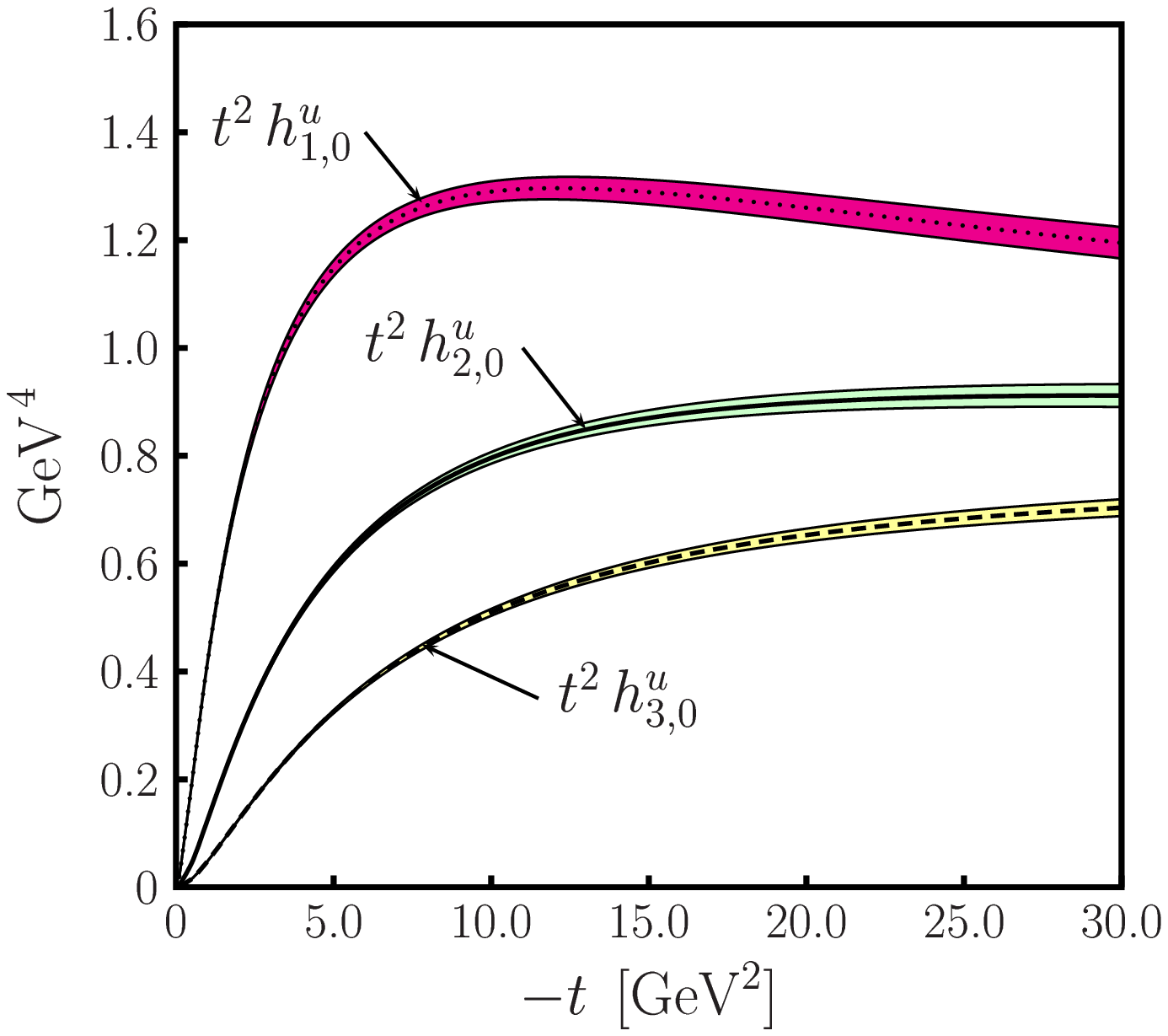}\hspace*{2em}
\includegraphics[width =.42\textwidth,
 bb = 103 338 447 677,clip=true]{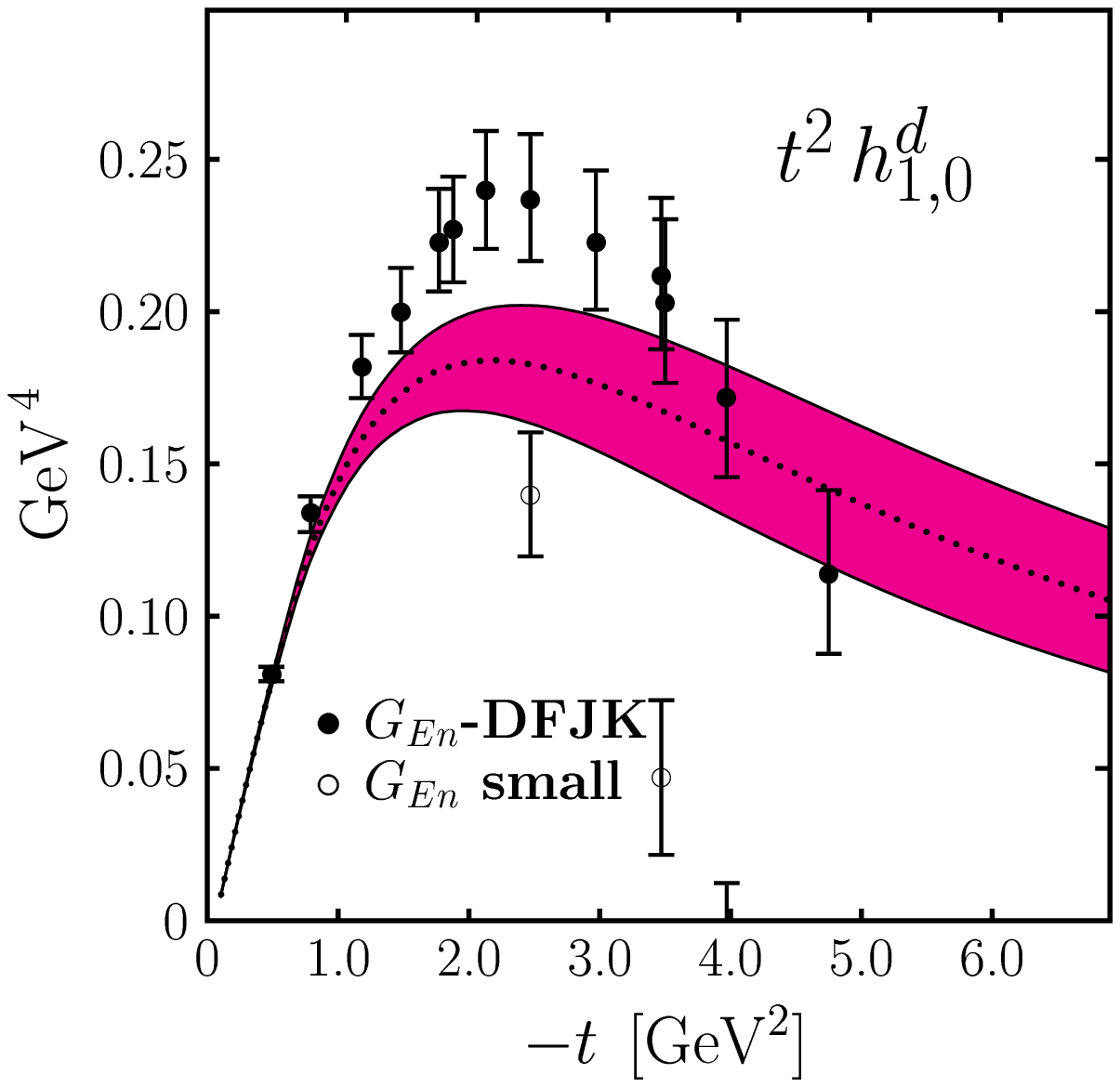}
\end{center}
\vspace*{-1.6em}
\caption{\label{fig:moment} Left: The first three moments
  of $H_v^u$ scaled by $t^2$. The error bands denote the
  parametric uncertainties. Right: The lowest moment of $H_v^d$ scaled
  by $t^2$. The shaded band represents the results obtained in Ref.\ 
  \ci{DFJK4}. Data are taken from \protect~\ci{brash,CLAS}.}
\end{figure}

Having the GPDs at disposal one can evaluate various moments, some of
them are displayed in Fig.\ \ref{fig:moment}. 
Comparison with recent results from lattice QCD~\ci{SESAM} reveals 
remarkable agreement of their $t$ dependencies given the uncertainties 
of the GPD analysis~\ci{DFJK4} and of the lattice calculations. It is 
to be stressed that the results presented in Refs.\ \ci{SESAM} are 
provided for a scenario in which the pion has a mass of 600 to 800
$\mev$. The extrapolation to the chiral limit is therefore problematic and 
may entail failures of the normalization of the moments.
An interesting property of the moments is that the $u$ and $d$ quark
contributions decrease with different rates at large $-t$. Because at 
large $-t$ the dominant contribution to the form factors comes from a 
narrow region of large $x$, see Fig.\ \ref{fig:gpdH}, one can use the 
large $x$ approximations
\be
q_v \sim (1-x)^{\beta_q}\,, \qquad f_q \sim A_q(1-x)^2\,,
\ee
and evaluate the sum rule \req{pn-sr} in the saddle point approximation. 
This leads to
\be
h^q_{1,0} \sim |t|^{-(1+\beta_q)/n}\,, \quad 
       1-x_s\= \left(\frac{n}{\beta_q}A_q|t|\right)^{-1/n}\,,
\label{power}
\ee
where $x_s$ is the position of the saddle point. For $n=2$ the saddle
point lies within the region where the GPD is large~\footnote{
For $n=1$ this is only the case for $-t$ larger than 
$30\,\gev^2$. The onset of the corresponding power behaviour of the
form factors occurs for larger $-t$.}
and using the the CTEQ values for $\beta_q$ ($\beta_u\simeq 3.4$ and
$\beta_d\simeq 5$ \ci{CTEQ}) one obtains a drop of the form
factor $h^u_{1,0}$ slightly faster than $t^{-2}$ while the $d$-quark
form factor falls as $t^{-3}$. Strengthened by the charge factor the
$u$-quark contribution dominates the proton's Dirac form
factor~\footnote{
This implies the ratio $F_1^n/F_1^p = e_d/e_u$ at large $-t$.}
for $-t$ larger than about $5\,\gev^2$, the $d$-quark contribution
amounts to less than $10\%$. This is what can be seen in Figs.\ 
\ref{fig:moment}. The power behavior bears resemblance to the 
Drell-Yan relation~\ci{DY}. In fact the common underlying dynamics is
the Feynman mechanism~\footnote{
The Feynman mechanism applies in the soft region where $1-x\sim
\Lambda/\sqrt{-t}$ and the virtualities of the active partons are
$\sim \Lambda\sqrt{-t}$ ($\Lambda$ is a typical hadronic scale).}.
The Drell-Yan relation is, however, an asymptotic result 
($x\to 1$, $t\to -\infty$) which bases on the assumption of valence
Fock state dominance, i.e.\ on the absence of sea quarks. The
different powers for $u$ and $d$ quarks signal that the asymptotic
region where the dimensional counting rules apply, has not yet been
reached. 
\begin{figure}
\begin{center}
\includegraphics[width =.40\textwidth,
 bb = 154 355 479 693,clip=true]{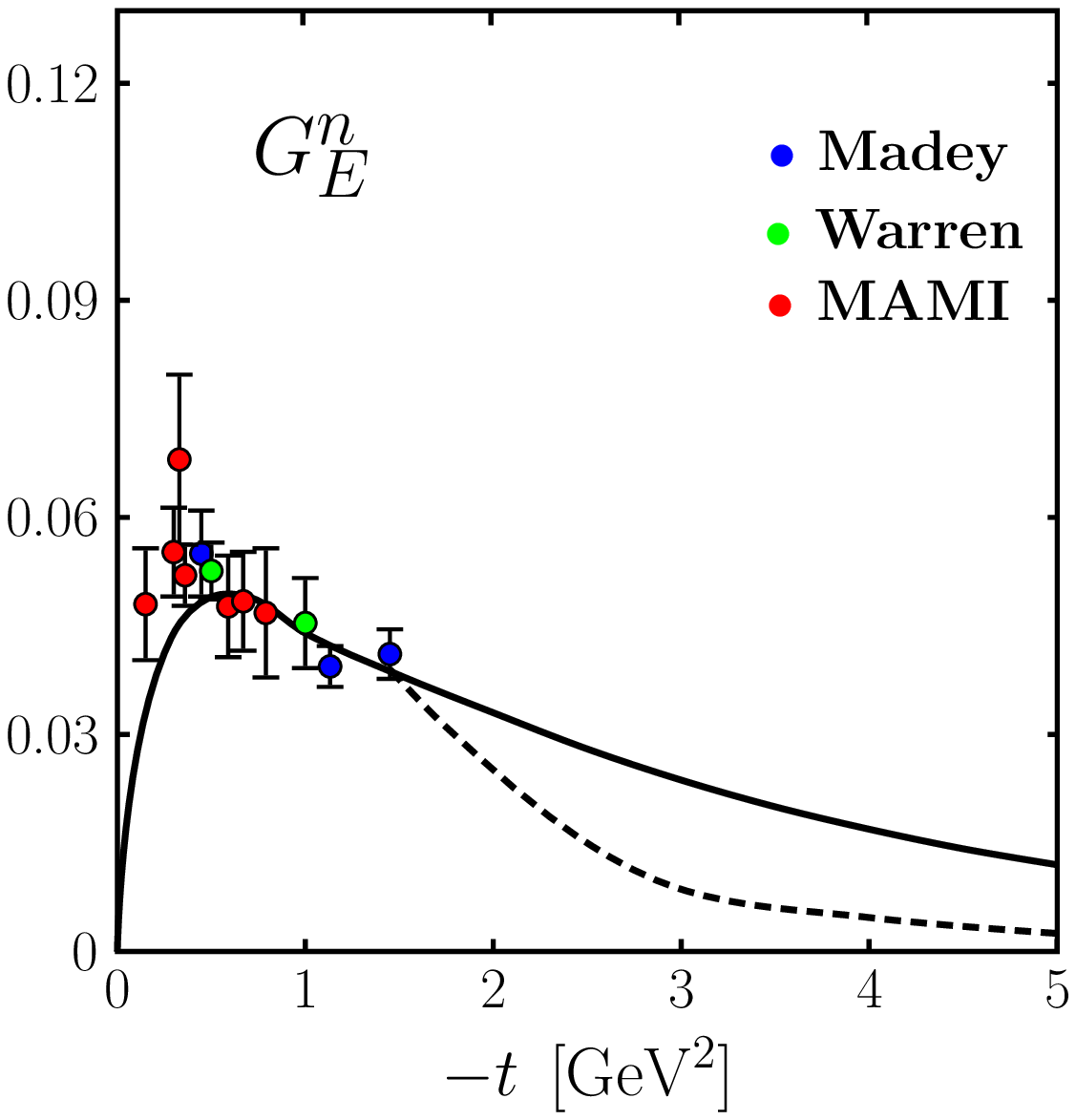} \hspace*{2em} 
\includegraphics[width=0.4\textwidth,bb=126 390 404 675,clip=true]
{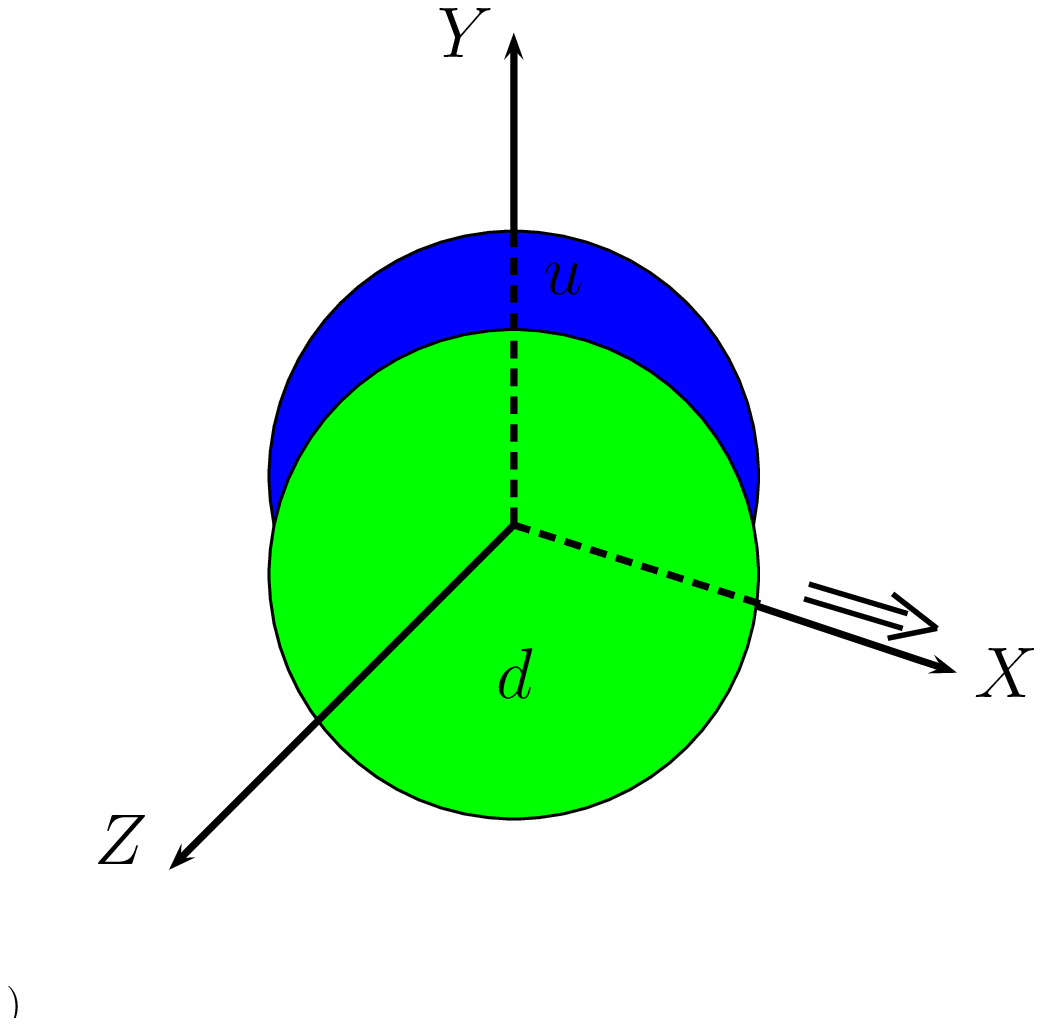}
\end{center}
\vspace*{-1.6em}
\caption{\label{fig:neutron} Left: The electric form factor of the
  neutron versus $t$ \protect~\ci{gen}. The lines represent different
  extrapolations, see text. Right: Sketch of flavor segregation in a
 transversally polarized proton. }
\end{figure}

One may object to the different powers that they are merely a
consequence of the chosen parameterization \req{ansatz},
\req{profile}. However, this is likely not the case. The moments
$h_{1,0}^q$ can directly be extracted from the form factor
data~\ci{brash}. The experimental results for the $d$-quark moment are
shown in Fig.\ \ref{fig:moment}. The sharp drop of it for increasing
$-t$ is clearly visible. The small deviations between data and the
moment obtained in \ci{DFJK4} are caused by the very precise but still
preliminary CLAS data~\ci{CLAS} for $G_M^n$. These data have been utilized
in the extraction of the moments $h_{1,0}^q$ but not in the analysis
in Ref.\ \ci{DFJK4}. Worst measured of the four form factors is the
electric one of the neutron. The recent data~\ci{gen} on it are 
shown in Fig.\ \ref{fig:neutron}. Above $1.5\,\gev^2$ no data is
available as yet and extrapolations to larger $-t$ are to be used. The 
solid line has been used in Ref.\ \ci{DFJK4} and in the determination 
of $h_{1,0}^q$ from data. If the electric form factor is smaller
than that extrapolation (e.g.\ the dashed line) $h_{1,0}^d$ falls even
faster. Thus, except $G_E^n$ is much larger than expected, the rapid
decrease of $h_{1,0}^d$ seems to be an experimental fact. Future JLab
data on $G_E^n$ will settle this question.  

A combination of the second moments of $H$ and $E$ at
$t=0$ is Ji's sum rule \cite{ji97} which allows for an evaluation of 
the valence quark contribution to the orbital angular momentum the quarks 
inside the proton carry
\be
\langle L_v^q\rangle \= \frac12\,\int_0^1 dx
       \Big[xE_v^q(x,t=0) + x q_v(x)- \Delta q_v(x)\Big]\,.
\ee
The analysis of the GPDs leads to 
\be
\langle L_v^u \rangle \= -(0.24 - 0.27)\,,\quad \langle L_v^d \rangle
\= 0.15 - 0.19\,,
\ee
for the valence quark contributions to the orbital angular momentum at
a scale of $\mu=2\,\gev$.

While the parton distributions only provide information on the
longitudinal distribution of quarks inside the nucleon, GPDs also give 
access to the transverse position distributions of partons within the
proton. Thus, the Fourier transform of $H$ 
\be
q_v(x,{\bf b}) \= \int \frac{d^2{\bf \Delta}}{(2\pi)^2}\, 
                e^{-i{\bf b}\cdot{\bf \Delta}}\, H_v^q(x,t=-\Delta^2)\,,
\ee  
gives the probability of finding a valence quark with longitudinal momentum
fraction $x$ and impact parameter ${\bf b}=(b^X,b^Y)$ as seen in a frame in
which the proton moves rapidly in the $Z$ direction. Together with the
analogous Fourier transform of $E_v^q(x,t)$ one can form the
combination ($m_p$ being the mass of the proton)
\be
q_v^X(x,{\bf b}) \= q_v(x,{\bf b}) - \frac{\;b^Y}{m_p}\,
\frac{\partial}{\partial {\bf b}^2}\, e_v^q(x,{\bf b})\,,
\ee
which gives the probability to find an unpolarized valence quark with
momentum fraction $x$ and impact parameter ${\bf b}$ in a
proton that moves rapidly along the $Z$ direction and is polarized
along the $X$ direction \ci{burk}. As shown in Ref.\ \ci{DFJK4}, for
small $x$ one observes a very broad distribution while at large $x$ 
it becomes more focused on the center of momentum defined by 
$\sum_i x_i {\bf b}_i=0$ ($\sum_i x_i=1$). In a proton that is
polarized in the $X$ direction the symmetry around the $Z$ axis is
lost and the center of the density is shifted in the $Y$ direction
away from the center of momentum, downward for $d$ quarks and upward 
for $u$ ones. In other words, a polarization of the proton induces a flavor 
segregation in the direction orthogonal to the direction of the 
polarization and the proton momentum, see Fig.\
\ref{fig:neutron}. This effect may be responsible for certain
asymmetries as, for instance, that one observed in $p\uparrow p\to \pi^\pm X$.        
\section{Wide-angle scattering}
\begin{figure}[t]
\begin{center}
\includegraphics[width=.47\textwidth,
  bb= 50 107 387 430,clip=true]{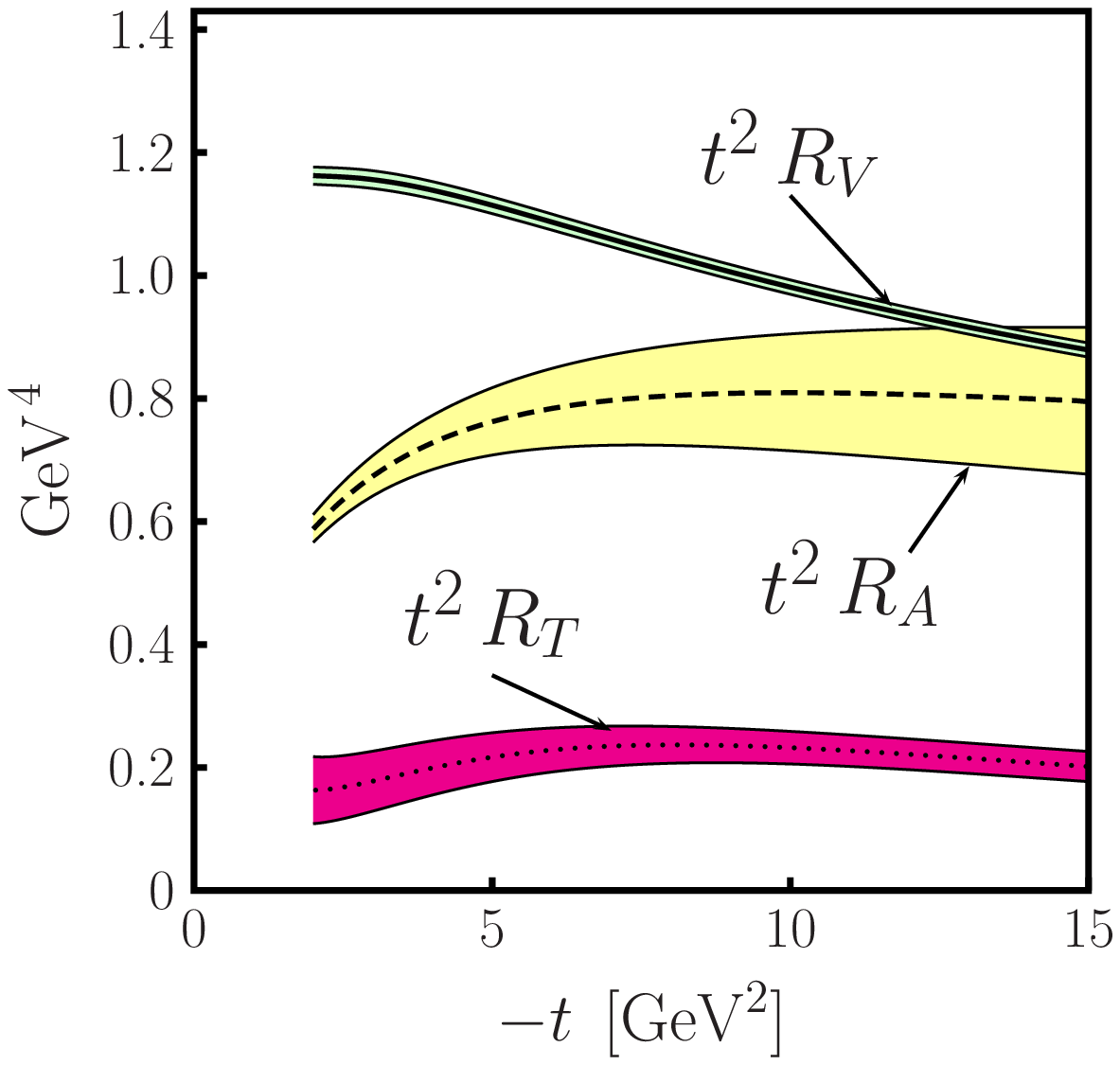}
\hspace*{2em}
\includegraphics[width=.45\textwidth, bb= 149 289 569 715,clip=true]
{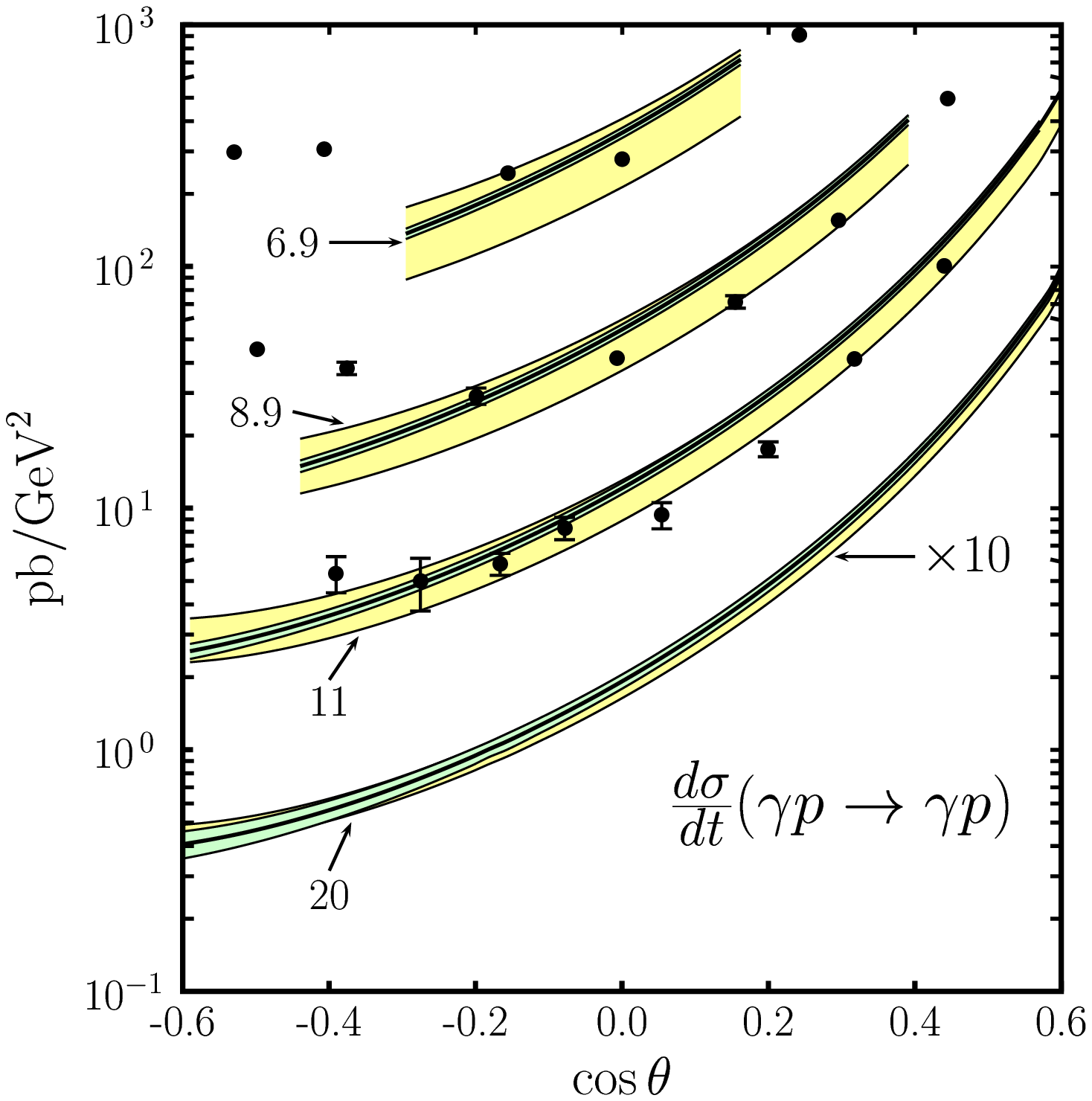}
\end{center}
\vspace*{-1.6em}
\caption{Left: The Compton form factors, scaled by $t^2$, evaluated from 
the GPDs determined in Ref.\ \protect\ci{DFJK4}. The bands represent
the parametric uncertainties of the form factors. Right: The Compton
cross section versus the c.m.s.\ scattering angle $\theta$ at 
$s=6.9,\,8.9,\,11$ and $20\,\gev^2$. The error bands are explained
in the text. Data are taken from Ref.\ \ci{nathan}.}
\label{fig:compton}
\end{figure}

As mentioned previously the analysis of the GPDs gives
insight in the transverse distribution of quarks inside the proton. 
However, there is more in it. With the $\xi=0$ GPDs at hand one can 
predict hard wide-angle exclusive reactions like Compton
scattering off protons or meson photo- and electroproduction. For
these reactions one can work in a so-called symmetrical frame where the
skewness is zero. It has been argued \ci{DFJK1,rad98} 
that, for large Mandelstam variables ($s, \;-t, \; -u \gg \Lambda^2$), the 
amplitudes for these reactions factorize into a hard partonic subprocess, 
e.g.\ Compton scattering off quarks, and in form factors representing 
$1/x$-moments of zero-skewness GPDS (see Fig.\ \ref{fig:compton}). 
For Compton scattering these form factors read 
\ba
R_V(t) &\simeq&\sum_{q=u,d} e_q^2 \int_{0}^1 \frac{dx}{x}\, H^q_v(x,t)\,, \nn\\
R_A(t) &\simeq& \sum_{q=u,d} e_q^2 \int_{0}^1 \frac{dx}{x}\, 
\widetilde{H}^q_v(x,t)\,, \nn\\[0.2em]
R_T(t) &\simeq&\sum_{q=u,d} e_q^2 \int_{0}^1 \frac{dx}{x}\, E^q_v(x,t)\,.
\label{Compton-formfactors}
\ea
In these expressions contributions from sea quarks have been 
neglected~\footnote{
An estimate of the sea quark contributions may be obtained by using 
the ansatz \req{ansatz} with the same profile function for the sea 
quarks as for the valence ones but replacing the valence quark density
with the CTEQ \ci{CTEQ} antiquark ones. The so estimated contributions 
are indeed small.}.
An analogous pseudoscalar form factor related to the GPD $\widetilde{E}$,
decouples in the symmetric frame. Numerical results for the Compton
form factors are shown in Fig.\ \ref{fig:compton}. To leading-order 
of pQCD the Compton cross section reads\ci{DFJK1,HKM} 
\ba
\frac{d\sigma}{dt} &=& \frac{d\hat{\sigma}}{dt} \left\{ \frac12\, \Big[
R_V^2(t)\, + \frac{-t}{4m^2} R_T^2(t) + R_A^2(t)\Big] \right.\nn\\
&&\hspace*{-0.5cm}\left.  - \frac{u s}{s^2+u^2}\,
\Big[R_V^2(t)\,+ \frac{-t}{4m^2} R_T^2(t) - R_A^2(t)\Big]\right\}\,,
\label{dsdt}
\ea
where $d\hat{\sigma}/dt$ is the Klein-Nishina cross section for
Compton scattering off massless, point-like spin-1/2 particles of
charge unity. Inserting the Compton form factors \req{Compton-formfactors} 
into Eqs.\ \req{dsdt}, one can predict the Compton cross section in the 
wide-angle region. The results for sample values of $s$ are shown 
in Fig.\ \ref{fig:compton} and compared to recent measurements from
JLab~\ci{nathan}. The inner bands of the predictions for $d\sigma/dt$
reflect the parametric errors of the form factors. The outer bands 
indicate the size of target mass corrections, see Ref.\ \cite{DFHK}. 
In order to comply with the kinematical requirements for handbag 
factorization, at least in a minimal fashion, predictions are only 
shown for $-t$ and $-u$ larger than about $2.5\, \gev^2$. Fair 
agreement between theory and experiment is to be seen. Next-to-leading 
order QCD corrections to the subprocess have been calculated in 
Ref.\ \ci{HKM}. They are not displayed in \req{dsdt} but taken into 
account in the numerical results. 

The handbag approach also applies to wide-angle photo- and
electroproduction of pseudoscalar and vector mesons~\ci{hanwen}. The 
amplitudes again factorize into a parton-level subprocess, $\gamma
q\to M q$ now, and form factors which represent $1/x$-moments of
GPDs. Their flavor composition reflects the valence quark content of
the produced meson. Therefore, the form factors which occur in 
photo-and electroproduction of pions and $\rho$ mesons, can also be
evaluated from the GPDs given in Ref.\ \ci{DFJK4}.

One may also consider the time-like process 
$\gamma\gamma\leftrightarrow p\bar{p}$ within the handbag approach
\ci{DKV}. Similar form factors as in the space-like region occur but
they are now functions of $s$ and represent moments of the $p\bar{p}$
distribution amplitudes, time-like versions of GPDs. With sufficient
data on the time-like electromagnetic form factors at disposal one may 
attempt a determination of the time-like GPDs. 
\section{Outlook}
Results from a first analysis of GPDs at zero skewness have been
reviewed. This analysis, has been performed in analogy to those of the usual
parton distributions, rests upon a physically motivated parameterization
of the GPDs fitted to the available nucleon 
form factor data. The analysis provides results on the valence-quark
GPDs $H$, $\widetilde{H}$ and $E$. The distribution of the quarks in
the impact parameter plane transverse to the direction of the nucleon's 
momentum can be evaluated from them. Polarizing the nucleon induces 
a flavor segregation in the direction orthogonal to the those of the 
nucleon's momentum and of its polarization. The average orbital angular 
momentum of the valence quarks can be estimated from the obtained
GPDs, too. Within the handbag approach the soft physics input to hard
wide-angle exclusive reactions is encoded in specific form factors
which represent $1/x$ moments of zero-skewness GPDs. Using the GPDs 
determined in Ref.\ \ci{DFJK4}, one can evaluate these form factors 
and, for instance, predict the wide-angle Compton cross section. The 
results are found to be in good agreement with experiment.

An analysis as that one performed in Ref.\ \ci{DFJK4} can only be
considered as a first attempt towards a determination of the GPDs. It
needs improvements in various aspects. High quality data on the form
factors at larger $t$ are required in order to stabilize the
parameterization. JLab will provide such data in the near future. As already 
mentioned CLAS \ci{CLAS} will come up with data on $G_M^n$ up to about 
$5\,\gev^2$. $G_E^n$ will be measured up to about $3.3\,\gev^2$ next
year and $G_E^p$ up to $9\,\gev^2$ in 2007. 
The upgraded Jlab will allow measurements of the nucleon form factors
up to about $13\,\gev^2$. Data on the axial form factor are needed in
order to improve our knowledge of $\widetilde{H}$.

Up to now only the sum rules \req{pn-sr}, \req{axial-sr} have been
utilized in the GPD analysis. As mentioned above, in this situation 
parameterizations of the GPDs are required with the consequence of
non-unique results. An alternative ansatz~\ci{DFJK4} is for instance 
\be
H_v^q(x,t) \= q_v(x)\Big[1-t f_q(x)/p\Big]^{-p}\,.
\label{ansatz-power}
\ee
Although reasonable fits to the form factors are obtained with 
\req{ansatz-power} for $p\gsim 2.5$, it is physically less
appealing than the parameterization \req{ansatz}: the combination of 
Regge behavior at small $x$ and $t$ with the dynamics of the Feynman 
mechanism is lost. The resulting GPDs have a broader shape and 
$H(x=0,t)$ remains finite. Thus, small $x$ also contribute to the 
large-$t$ form factors. Higher moments are needed in order to lessen 
the dependence on the chosen parameterization for the GPDs. Such
moments can be provided by lattice gauge theories. The present lattice
results \ci{SESAM} are however calculated in scenarios where the
pion is heavy (typically $600 - 800\,\mev$) and the extrapolation
to the chiral limit is uncertain. Obviously such
results are inappropriate for use in a GPD analysis. In a few years
the quality of the moments from lattice gauge theories may
suffice. It is also tempting to use the data on wide-angle Compton
scattering~\ci{nathan} which imply information on $1/x$ moments of 
the GPDs. However, even at the highest measured energy, $s=11\,\gev^2$, the 
Compton cross section may still be contaminated by power corrections 
rendering its use in a GPD analysis dubious. For data at, say, 
$s=20\,\gev^2$ the situation may be different. Instead of a
parameterization one may use the maximum entropy method \ci{faccioli} 
for the extraction of the GPDs from the sum rules. This method has not 
yet been utilized for this purpose.

\begin{figure}[t]
\begin{center}
\includegraphics[width=.50\textwidth, 
  bb= 101 210 569 702,clip=true]{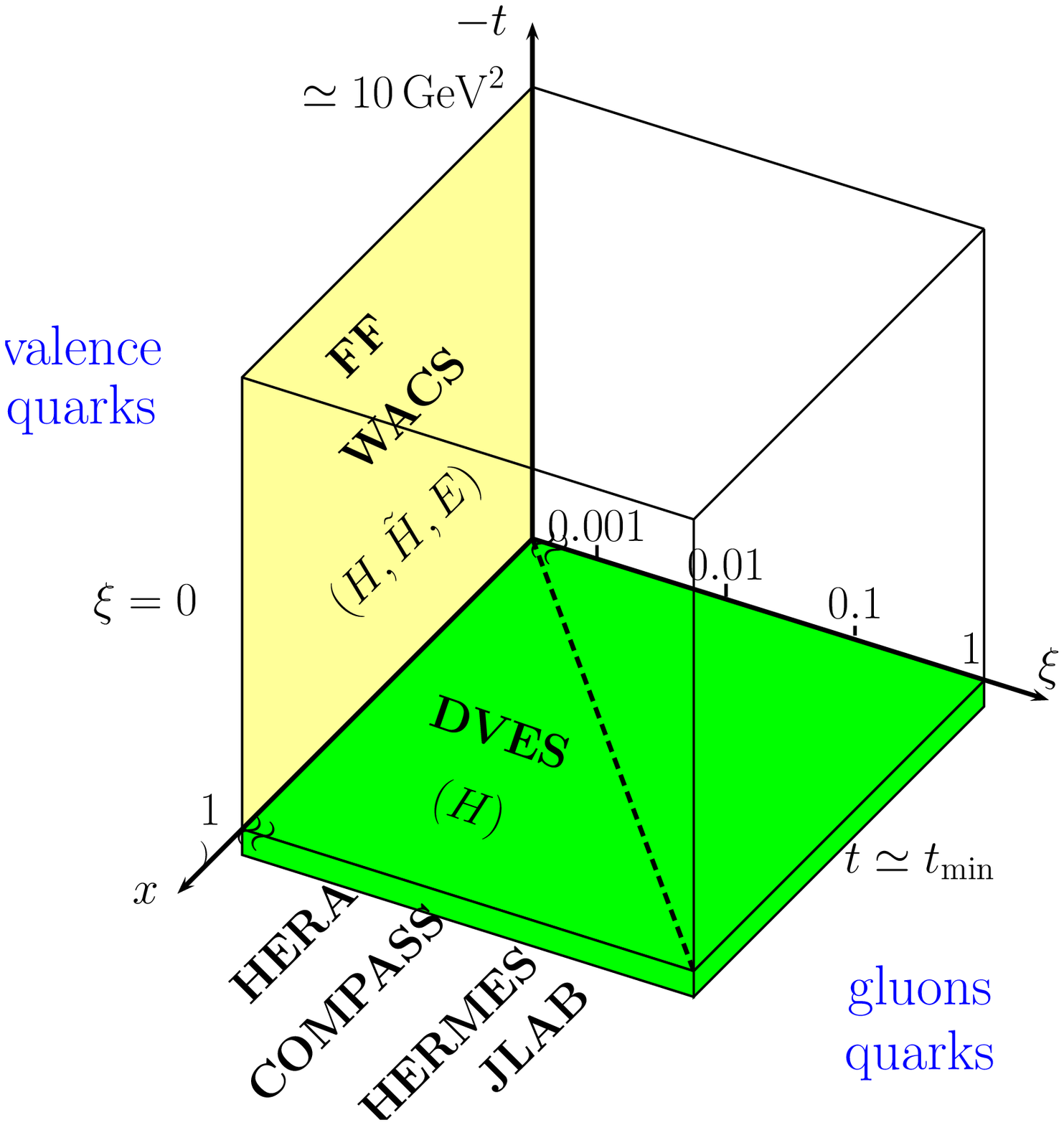}
\end{center}
\vspace*{-1.6em}
\caption{\label{fig:GPD} Experimental information on GPDs.}
\end{figure}

Up to now only the zero-skewness GPDs have been determined. Although
the sum rules \req{pn-sr}, \req{axial-sr} are valid at all values of the 
skewness it seems a hopeless task to extract functions of three
variables from them. Additional information is demanded in the case of
$\xi \neq 0$ and will be provided by deeply virtual exclusive
scattering (DVES). The skewness is related to Bjorken-$x$ in DVES and, 
hence, non-zero. At large $Q^2$ but small $t$ the amplitudes for DVES
are given by a convolution of GPDs and appropriate subprocess
amplitudes. The convolution possesses poles at $x =\pm \xi$.
For Compton scattering the pole terms can directly be isolated in the interference
region with the Bethe-Heitler process and will lead to a nearly 
model-independent determination of $H(\xi,\xi,t)$. For $x\neq \xi$  
parameterizations of the GPDs will most likely be needed. At present 
a number of experiments on DVES are running (HERA, COMPASS, HERMES,
JLab). In Fig.\ \ref{fig:GPD} kinematical regions are shown as  
shaded areas in which information on GPDs is currently accumulated. 
A lot of work is ahead of us before we can say that we have a fair
knowledge on the GPDs.
\section*{Acknowledgments} It is a pleasure to thank Anatoly Efremov
and the other members of the organizing committee for organizing this 
interesting
meeting and for the hospitality extended to him at Dubna. This work
has been supported in part by the Heisenberg-Landau program.

\end{document}